\documentclass[preprint,showpacs,amsmath,amssymb,aps,prd]{revtex4}
\usepackage{graphicx}

\usepackage{amsmath}

 \newcommand{\ds}{\displaystyle}

\begin{document}

\title{Geometric sigma model of the Universe}

\author{Milovan Vasili\'c} \email{mvasilic@ipb.ac.rs}
\affiliation{Institute of Physics, University of Belgrade, P.O.Box 57,
11001 Belgrade, Serbia}

\date{\today}

\begin{abstract}
The purpose of this work is to demonstrate how an arbitrarily chosen
background of the Universe can be made a solution of a simple
geometric sigma model. Geometric sigma models are purely geometric
theories in which spacetime coordinates are seen as scalar fields
coupled to gravity. Although they look like ordinary sigma models,
they have the peculiarity that their complete matter content can be
gauged away. The remaining geometric theory possesses a background
solution that is predefined in the process of constructing the theory.
The fact that background configuration is specified in advance is
another peculiarity of geometric sigma models. In this paper, I
construct geometric sigma models based on different background
geometries of the Universe. Whatever background geometry is chosen,
the dynamics of its small perturbations is shown to posses a generic
classical stability. This way, any freely chosen background
metric is made a stable solution of a simple model. Three particular
models of the Universe are considered as examples of how this is done
in practice.
\end{abstract}

\pacs{04.50.Kd, 98.80.Jk}

\maketitle

\section{Introduction}\label{Sec1}

The latest astronomical observations have given a substantial boost to
the development of modern cosmology \cite{1, 2, 3, 4, 5, 6, 7, 8, 9,
10}. In particular, the accelerating expansion of the Universe has
drawn much attention. The early time acceleration is widely known as
{\it inflation}, while the late time acceleration is usually referred
to as the epoch of {\it dark energy} \cite{11, 12, 13, 14}. Presently,
the $\Lambda$CDM model, in which the cosmological constant $\Lambda$
plays the role of dark energy, is accepted as a standard cosmological
model. There is an extensive literature on other forms of dark energy,
too \cite{21, 22, 23, 24, 25, 26, 27, 28, 29, 30, 31}. All in all, the
number of dark energy models that can be found in literature is
enormous. The same holds for the inflationary models that have been
constructed over the years.

In this paper, I shall describe the procedure which associates an
action functional with an arbitrarily chosen background geometry of
the Universe. Precisely, any desirable geometry of the Universe is
made a solution of a particular geometric sigma model. Geometric
sigma models are theories that possess two distinctive properties.
First, their complete matter content can be gauged away. Second, any
predefined geometry can be made a solution of a particular model.
These models have first been proposed in \cite{32} in the context of
fermionic excitations of flat geometry. Here, I use them for modeling
the dynamics of the Universe. To be more accurate, only geometry and
dark energy are considered in this approach. The inclusion of
ordinary matter is discussed separately.

The results obtained in this paper are summarized as follows. First, a
class of purely geometric dark energy models has been constructed.
Every particular model is defined as a geometric sigma model
associated with a spatially flat, homogeneous and isotropic geometry.
This way, an arbitrarily chosen geometry of this kind becomes a
background solution of a particular geometric sigma model. Ultimately,
one is provided with the class of dark energy models parametrized by
their background geometries. The inflation and the late time
acceleration have purely geometric origin. It is important to
emphasize that, while the background metric can be chosen arbitrarily,
the physics of its small perturbations can not. In fact, the
background metrics just parametrize geometric sigma models, very much
the same as inflaton potentials parametrize the inflationary models.

The second result concerns the linear stability of the background
solution in geometric sigma models. It has been proven true for almost
all background geometries. Precisely, the stability is guaranteed up
to the existence of critical moments, where the perturbations may
diverge. There, however, the linear analysis fails, and should be
corrected by the inclusion of interaction terms.

Finally, I have analyzed geometric sigma models coupled to ordinary
matter. It has been shown that matter fields do not compromise the
vacuum stability established earlier. In the case of minimal coupling
to the metric, the linear stability of matter itself has been proven.

The layout of the paper is as follows. In Sec. \ref{Sec2}, the
construction of geometric sigma models, as defined in \cite{32}, is
recapitulated and subsequently applied to spatially flat, homogeneous
and isotropic geometries. As a result, a class of action functionals
of the Universe is obtained. Each of these action functionals
possesses a nontrivial background solution that describes the
background geometry of a particular Universe. In Sec. \ref{Sec3}, the
dynamics of small perturbations of these nontrivial backgrounds is
examined. In Sec. \ref{Sec4}, the background solutions are proven
stable for almost all spatially flat, homogeneous and isotropic
geometries. In Sec. \ref{Sec4.5}, geometric sigma modes are coupled
to ordinary matter. It is shown that matter fields preserve the
results obtained in the absence of matter. In Sec. \ref{Sec5}, the
examples of inflationary and bouncing Universes are used to
demonstrate how geometric sigma models are constructed in practice.
Sec. \ref{Sec6} is devoted to concluding remarks.

My conventions are as follows. The indexes $\mu$, $\nu$, ... and $i$,
$j$, ... from the middle of the alphabet take values $0,1,2,3$. The
indexes $\alpha$, $\beta$, ... and $a$, $b$, ... from the beginning
of the alphabet take values $1,2,3$. The spacetime coordinates are
denoted by $x^{\mu}$, the ordinary differentiation uses comma
($X_{,\,\mu} \equiv \partial_{\mu} X$), and the covariant
differentiation uses semicolon ($X_{;\,\mu}\equiv \nabla_{\mu}X$).
The repeated indexes denote summation: $X_{\alpha\alpha} \equiv
X_{11}+X_{22}+X_{33}$. The signature of the $4$-metric $g_{\mu\nu}$
is $(-,+,+,+)$, and the curvature tensor is defined as
$R^{\mu}{}_{\nu\lambda\rho} \equiv \partial_{\lambda}
\Gamma^{\mu}{}_{\nu\rho} - \partial_{\rho}
\Gamma^{\mu}{}_{\nu\lambda} + \Gamma^{\mu}{}_{\sigma\lambda}
\Gamma^{\sigma}{}_{\nu\rho} - \Gamma^{\mu}{}_{\sigma\rho}
\Gamma^{\sigma}{}_{\nu\lambda}$.

\section{Geometric sigma models}\label{Sec2}

The construction of a geometric sigma model begins with specifying a
spacetime metric. I shall denote it with $g_{\mu\nu}^{(o)}(x)$. The
metric $g_{\mu\nu}^{(o)}$ is freely chosen, and the coordinates
$x^{\mu}$ are fully fixed. As a consequence, the functions
$g_{\mu\nu}^{(o)}(x)$ are completely determined. In the next step,
the corresponding Ricci tensor $R_{\mu\nu}^{(o)}(x)$ is calculated,
and the following Einstein like equation is postulated:
\begin{equation} \label{1}
R_{\mu\nu} = R_{\mu\nu}^{(o)}(x) \,.
\end{equation}
Obviously, the metric $g_{\mu\nu}^{(o)}$ is a solution of the equation
(\ref{1}). Its non-zero right hand side {\it defines} matter content
of the theory. The equation (\ref{1}) is an example of how an
arbitrarily chosen metric can be made a solution of a simple model.

The equation (\ref{1}) obviously lacks general covariance. To
covariantize it, I introduce a new set of coordinates
$\phi^i=\phi^i(x)$. In terms of these new coordinates, the equation
(\ref{1}) takes the form
\begin{equation} \label{2}
R_{\mu\nu} = H_{ij}(\phi) \phi^i_{,\mu} \phi^j_{,\nu} \,,
\end{equation}
where the functions $H_{ij}(\phi)$ are defined through
\begin{equation} \label{3}
H_{ij}(\phi) \equiv R_{ij}^{(o)}(\phi) \,.
\end{equation}
In other words, the ten functions $H_{ij}(\phi)$ are obtained by
replacing $x$ with $\phi$ in ten components of the Ricci tensor
$R_{\mu\nu}^{(o)}(x)$. The equation (\ref{2}) is generally covariant
once the new coordinates $\phi^i$ are seen as scalar functions of the
old coordinates $x^{\mu}$. If the new coordinates are chosen to
coincide with the old ones, $\phi^i(x)\equiv \delta^i_{\mu} x^{\mu}$,
the covariant equation (\ref{2}) is brought back to the non-covariant
form (\ref{1}). In what follows, the shorthand notation
$\delta^i_{\mu} x^{\mu}\equiv x^i$ will be used.

The equation (\ref{2}) has the form of the Einstein's
equation in which four scalar fields $\phi^i(x)$ of some nonlinear
sigma model are coupled to gravity. The "matter field equations" are
obtained by utilizing the Bianchi identities $(R^{\mu\nu} -\frac{1}{2}
g^{\mu\nu} R)_{;\nu}\equiv 0$. If the condition $\det \phi^i_{,\mu}
\neq 0$ is fulfilled, one obtains
\begin{equation} \label{4}
H_{ij}\nabla^2\phi^j + {1\over 2}\left({{\partial H_{ij}}\over
{\partial \phi^k}}-{{\partial H_{jk}}\over {\partial \phi^i}}+
{{\partial H_{ki}}\over {\partial \phi^j}}\right)\phi^j_{,\mu}
\phi^{k,\mu} = 0 \, .
\end{equation}
The equation (\ref{4}) is not an independent equation, as it follows
from (\ref{2}) and the Bianchi identities. It is straightforward to
verify that the equations (\ref{2}) and (\ref{4}) can be derived from
the action functional
\begin{equation} \label{5}
I_g = \frac{1}{2\kappa}\int d^4x\sqrt{-g}\left[ R -
H_{ij}(\phi)\phi^i_{,\mu}\phi^{j,\mu} \right] .
\end{equation}
Here, the target metric $H_{ij}(\phi)$ is not an independent
coefficient of the model. Instead, it is constructed out of the
background metric $g_{\mu\nu}^{(o)}$, through its defining relation
(\ref{3}). This way, an action functional is associated with every
freely chosen background metric. This action functional describes a
nonlinear sigma model coupled to gravity, and possesses the nontrivial
(let me call it {\it vacuum}) solution
\begin{equation} \label{6}
 \phi^i = x^i \,, \qquad
g_{\mu\nu} = g_{\mu\nu}^{(o)}\,.
\end{equation}
Indeed, the equation (\ref{2}) with the target metric (\ref{3}) is
trivially satisfied if the scalars $\phi^i$ and the metric $g_{\mu\nu}$
are given by (\ref{6}). Being a direct consequence of (\ref{2}), so is
the equation (\ref{4}).

The physics of small perturbations of the vacuum (\ref{6}) does not
violate the condition $\det \phi^i_{,\mu} \neq 0$, which enables one
to interpret the scalars $\phi^i$ as spacetime coordinates. If this is
the case, one is allowed to fix the gauge $\phi^i(x)=x^i$, which
brings us back to the geometric equation (\ref{1}). One should have in
mind, however, that the action (\ref{5}) has a non-geometric sector,
too. It is characterized by $\det \phi^i_{,\mu} = 0$, and includes
trivial vacuum solutions such as $\phi^i = \rm{const}$. In what
follows, I shall restrict my considerations to purely geometric
dynamics of small perturbations of the vacuum (\ref{6}).

Before I continue, let me note that the equation (\ref{1}) is not the
unique geometric equation that allows the solution $g_{\mu\nu} =
g_{\mu\nu}^{(o)}$. A simple generalization of this equation can be
obtained by adding terms proportional to $g_{\mu\nu} -
g_{\mu\nu}^{(o)}$. The simplest choice is the equation
\begin{equation} \label{7}
R_{\mu\nu} = R_{\mu\nu}^{(o)}(x) + \frac{1}{2} V(x)
\left( g_{\mu\nu} - g_{\mu\nu}^{(o)} \right) \,.
\end{equation}
It defines a class of geometric theories parametrized by metrics
$g_{\mu\nu}^{(o)}$, and potentials $V$. The covariantization of the
non-covariant equation (\ref{7}) ultimately leads to the action
functional
\begin{equation} \label{8}
I_g = \frac{1}{2\kappa}\int d^4x\sqrt{-g}\left[ R -
F_{ij}(\phi)\phi^i_{,\mu}\phi^{j,\mu} - V(\phi) \right] \,,
\end{equation}
where the target metric $F_{ij}(\phi)$ is defined by
\begin{equation} \label{9}
F_{ij}(x)\equiv R_{ij}^{(o)}(x) - \frac{1}{2} V(x)
g_{ij}^{(o)}(x) \,.
\end{equation}
The class of theories defined by (\ref{8}) possesses the vacuum
solution (\ref{6}) for any choice of the potential $V(\phi)$. The
physics of small perturbations of this vacuum allows the gauge
condition $\phi^i = x^i$, which brigs us back to the geometric
equation (\ref{7}).

In what follows, I shall associate a geometric sigma model with a
given vacuum geometry of the Universe. Let me choose a spatially
flat, homogeneous and isotropic metric $g_{\mu\nu}^{(o)}$, defined by
\begin{equation} \label{10}
ds^2 = -dt^2 + a^2(t)\left( dx^2 + dy^2 + dz^2 \right) \,.
\end{equation}
The corresponding Ricci tensor is calculated straightforwardly. One
obtains
$$
R_{00}^{(o)} = -3\,\frac{\ddot a}{a} \,,   \quad
R_{0\alpha}^{(o)} = 0 \,,                \quad
R_{\alpha\beta}^{(o)} = \left(a\ddot a +
2\dot a^2 \right)\delta_{\alpha\beta} \,,
$$
where "dot" denotes time derivative. Now, I am ready to construct the
target metric $F_{ij}(\phi)$. In this paper, I choose the {\it
simplest model} allowed by (\ref{9}). It is obtained by noticing that
the target metric $F_{ij}$ can significantly be simplified by a
proper choice of the potential $V$. Indeed, if the potential is
chosen to have the form
\begin{equation} \label{12}
V(t) = 2\left(2\,\frac{\dot a^2}{a^2} + \frac{\ddot a}{a}\right) ,
\end{equation}
the component $F_{00}$ remains the only nonzero component of the
target metric. Precisely, one obtains
\begin{equation} \label{13}
F(t) = 2\left(\frac{\dot a^2}{a^2} - \frac{\ddot a}{a}\right) ,
\end{equation}
where the identification $F_{00}\equiv F$ is introduced for
convenience. As a consequence, $\phi^0$ is the only scalar field that
enters the action functional (\ref{8}). Let me simplify the notation
by using the identification $\phi^0\equiv \phi$. The action (\ref{8})
then reduces to
\begin{equation} \label{14}
I_g = \frac{1}{2\kappa}\int d^4x\sqrt{-g}\left[ R -
F(\phi)\phi^{,\mu}\phi_{,\mu} - V(\phi) \right] .
\end{equation}
It governs the dynamics of gravity coupled to a scalar field, and
possesses the vacuum solution
\begin{equation} \label{15}
\phi = t \,, \qquad g_{\mu\nu}=g^{(o)}_{\mu\nu}\,.
\end{equation}
The precise form of the target metric $F(\phi)$ and the potential
$V(\phi)$ is determined once the function $a(t)$ is specified. It
should be noted that it is only the background geometry
$g^{(o)}_{\mu\nu}$ that is freely chosen. The dynamics of metric
perturbations is governed by the corresponding action functional. The
class of action functionals (\ref{14}) represents a collection of
dark energy models parametrized by the scale factors $a(t)$. 

Similar attempts to derive a dark energy model out of the given scale
factor already exist in literature. Take, for example, references
\cite{35, 36, 37, 38, 39, 40, 41, 42, 43, 44}. There, scalar field
dark energy models have been constructed to mimic holographic dark
energy. Every particular model is built with a separate effort to solve
a particular problem. The procedure described in this paper, however,
is the first systematic approach of the kind. It gives a precise
prescription of how to construct the target metric and the potential
of a stable dark energy model. As we shall see later, the generic
classical stability is guaranteed for nearly any chosen background.
Some specific models are considered in Sec. \ref{Sec5}.

It should be noted that no ordinary matter has been considered so far.
Luckily, the inclusion of matter fields does not compromise the basic
predictions of dark energy models (\ref{14}). This will be
demonstrated later in Sec. \ref{Sec4.5}. Besides, the prevailing form
of matter in the Universe is believed to be the dark energy. Thus, the
class of dark energy models (\ref{14}) can roughly be interpreted as
zero approximation of more realistic cosmologies.

The standard physical requirements that ensure the absence of ghosts
and tachions restrain the target metric $F(\phi)$ to be positively
definite, and the potential $V(\phi)$ to be bounded from below. These
restrictions, however, refer to trivial vacuums. Precisely, the
positive definiteness of $F(\phi)$, and the fact that $\phi=\phi_0$
is a minimum of the potential $V(\phi)$ ensure stability of the
vacuum $\phi=\phi_0$, $g_{\mu\nu}=\eta_{\mu\nu}$. In this paper,
however, the vacuum of interest is the nontrivial vacuum (\ref{15}).
Its stability is not guaranteed by the above physical requirements,
and I am led to check it by direct calculation.

\section{Dynamics of small perturbations}\label{Sec3}

In this section, I shall examine the dynamics of small perturbations
of the vacuum (\ref{15}), as governed by the action functional
(\ref{14}). The infinitesimal change of coordinates $x^{\mu} \to
x^{\mu} + \xi^{\mu}(x)$ leaves this action invariant, and allows the
gauge fixing $\phi = t$. In this gauge, the matter field equation is
identically satisfied, and we are left with the gravitational field
equation (\ref{7}). The residual diffeomorphisms are defined by the
constraint $\xi^0=0$.

The only variable in the gauge fixed theory is the metric
perturbation $h_{\mu\nu}$, defined by
$$
g_{\mu\nu} = g^{(o)}_{\mu\nu} + h_{\mu\nu}.
$$
With respect to the residual diffeomorphisms, it transforms as
$$
\begin{array}{rcl}
&&\ds\delta_0 h_{00} = 0 \,,                            \\
&&\delta_0 h_{0\alpha}=-a^2 \dot\xi_{\alpha} \,,        \\
&&\ds\delta_0 h_{\alpha\beta}=-a^2 \left( \xi_{\alpha,\beta}+
\xi_{\beta,\alpha} \right),
\end{array}
$$
where $\delta_0$ is the form variation, and $\xi_{\alpha} \equiv
\xi^{\alpha}$. It is seen that $h_{0\alpha}$ can also be gauged away.
The gauge condition
$$
h_{0\alpha} = 0
$$
restrains the gauge parameters to be functions of spatial
coordinates, only. Precisely, the residual gauge parameters are
defined by
$$
\xi^0 = 0 \,, \qquad \dot\xi^{\alpha} = 0 \,.
$$
In what follows, I shall demonstrate how the residual gauge symmetry
can further be fixed.

Let me first linearize the field equations (\ref{7}). After
cumbersome, but straightforward, calculation one obtains
%
\begin{subequations}\label{20}
\begin{equation}\label{20a}
\partial_0\left( \dot h_{\alpha\alpha} -
2\frac{\dot a}{a}\,h_{\alpha\alpha}\right) +
3a\dot a \,\dot h_{00} + 2\left(a\ddot a +
2\dot a^2\right)h_{00} + h_{00,\alpha\alpha} = 0 \,,
\end{equation}
\begin{equation}\label{20b}
\partial_{0}\left[\,\frac{1}{a^2}\left(h_{\alpha\beta,\beta}-
h_{\beta\beta,\alpha}\right)\right] -
2\frac{\dot a}{a}\,h_{00,\alpha} = 0 \,,
\end{equation}
\begin{equation}\label{20c}
\begin{array}{l}
\ds\ddot h_{\alpha\beta}-\frac{\dot a}{a}\dot h_{\alpha\beta}-
2\frac{\ddot a}{a}h_{\alpha\beta} + \frac{1}{a^2}
\left( h_{\alpha\gamma,\gamma\beta} +
h_{\beta\gamma,\gamma\alpha}-
h_{\alpha\beta,\gamma\gamma}-
h_{\gamma\gamma,\alpha\beta}\right)                            \\
\ds + \left[\frac{\dot a}{a}\,\dot h_{\gamma\gamma}-
2\frac{\dot a^2}{a^2}\, h_{\gamma\gamma} +
a\dot a\, \dot h_{00} + 2\left( a\ddot a + 2\dot a^2
\right)h_{00}\right]\delta_{\alpha\beta} +
h_{00,\alpha\beta} = 0 \,.                                      \\
\end{array}
\end{equation}
\end{subequations}
%
It is immediately seen that the equation (\ref{20b}) implies
$$
\partial_0 \left[ \frac{1}{a^2}
\left(h_{\alpha\gamma,\gamma\beta}-
h_{\beta\gamma,\gamma\alpha}\right)\right] = 0 \,,
$$
which tells us that the expression in square brackets does not depend
on time. As a consequence, this expression can be gauged away.
Indeed, its transformation law reads
$$
\delta_0 \left[ \frac{1}{a^2}
\left(h_{\alpha\gamma,\gamma\beta}-
h_{\beta\gamma,\gamma\alpha}\right)\right] =
\xi_{\alpha,\gamma\gamma\beta}-
\xi_{\beta,\gamma\gamma\alpha} \,.
$$
Both, the expression in square brackets and the residual gauge
parameters $\xi_{\alpha}$ are functions of spatial coordinates alone.
This allows the gauge fixing
\begin{equation}\label{23}
h_{\alpha\gamma,\gamma\beta}-
h_{\beta\gamma,\gamma\alpha} = 0 \,.
\end{equation}
In what follows, I shall simplify the analysis by the assumption that
metric perturbations are {\it spatially localized}. This means that
the perturbation $h_{\mu\nu}$ is assumed to decrease sufficiently
fast in spatial infinity. With this assumption, many expressions are
simplified. For example, the equation $X_{,\alpha}=0$ has the general
solution $X=X(t)$, but the adopted boundary conditions imply $X=0$.
Similarly, the equation $X_{,\alpha\alpha}=0$ has the unique solution
$X=0$. With this in mind, one easily determines the residual symmetry
after the gauge condition (\ref{23}) has been imposed. It is defined
by
$$
\xi_{\alpha} = \epsilon_{,\alpha} \,,
$$
where the new parameter $\epsilon$ is an arbitrary function of
spatial coordinates.

Let me now extract the divergence free parts of the variable
$h_{\alpha\beta}$. To this end, I use the decomposition
$$
h_{\alpha\beta} \equiv \tilde h_{\alpha\beta} +
\tilde h_{\alpha,\beta} + \tilde h_{\beta,\alpha} +
\tilde h_{,\alpha\beta} \,,
$$
where $\tilde h_{\alpha\beta}$ and $\tilde h_{\alpha}$ are, by
definition, divergence free ($\tilde h_{\alpha\beta,\beta} \equiv
\tilde h_{\alpha,\alpha} \equiv 0$). In what follows, all the
expressions will be rewritten in terms of the new variables $\tilde
h_{\alpha\beta}$, $\tilde h_{\alpha}$ and $\tilde h$. Let me start
with the gauge condition (\ref{23}). With the help of the adopted
boundary conditions, it is straightforward to verify that (\ref{23})
becomes
$$
\tilde h_{\alpha} = 0 \,.
$$
Now, I am ready to rewrite the equations (\ref{20}) in terms of the
remaining variables $h_{00}$, $\tilde h_{\alpha\beta}$ and $\tilde
h$. Let me first consider the scalar sector. As it turns out, there
are only three independent scalar equations. The first two are the
constraint equations
\begin{subequations}\label{27}
\begin{equation}\label{27a}
a\ddot a\left[\frac{1}{a^2}\tilde h_{\alpha\alpha} \right]_{,0} +
\frac{\dot a}{a} \left[ \frac{1}{a^2} \tilde h_{\alpha\alpha}
\right]_{,\beta\beta} =
\,2\dot a^2 \left[ \frac{1}{a^2}\tilde h\right]_{,0\beta\beta} \,,
\end{equation}
\begin{equation}\label{27b}
\left[ \frac{1}{a^2} \tilde h_{\alpha\alpha} \right]_{,0} +
2\frac{\dot a}{a}\, h_{00} = 0 \,.
\end{equation}
\end{subequations}
The equation (\ref{27b}) follows from (\ref{20b}), and (\ref{27a}) is
a linear combination of the trace and divergence of (\ref{20c}). It
is seen that $h_{00}$ is fully determined by other variables. Thus,
it carries no degrees of freedom. On the other hand, the variable
$\tilde h/a^2$ is determined up to a free function of the spatial
coordinates. This freedom, however, can readily be gauged away.
Indeed, the variable $\tilde h/a^2$ transforms as
$$
\delta_0 \left(\frac{1}{a^2}\tilde h\right) = -2\epsilon
$$
with respect to the residual symmetry of the model. As $\tilde
h_{\alpha\alpha}/a^2$ is gauge invariant, the residual parameter
$\epsilon = \epsilon(\vec x)$ is exactly what one needs to fix the
free integration function of (\ref{27a}). In summary, the constraint
equations (\ref{27}) tell us that neither $h_{00}$ nor $\tilde h/a^2$
carry physical degrees of freedom.

The third scalar equation is the equation (\ref{20a}), or
equivalently, the divergence of (\ref{20c}). When its coefficients
are expressed in terms of the Hubble parameter $H\equiv \dot a/a$, it
takes the form
%
\begin{equation}\label{28}
H\dot H\left[ \frac{1}{a^2}\tilde h_{\alpha\alpha}\right]_{,00} +
\left(3\dot H H^2 - 2\dot H^2 + H\ddot H \right)
\left[ \frac{1}{a^2}\tilde h_{\alpha\alpha} \right]_{,0} -
\frac{1}{a^2}H\dot H \left[ \frac{1}{a^2}
\tilde h_{\alpha\alpha}\right]_{,\beta\beta} = 0 \,.
\end{equation}
%
The equation (\ref{28}) governs the dynamics of the unique scalar
mode of the model.

Now, I am left with the traceless, divergence free part of the
equation (\ref{20c}), which governs the dynamics of the traceless,
divergence free part of $h_{\alpha\beta}$. The latter is defined by
$$
\hat h_{\alpha\beta} \equiv \tilde h_{\alpha\beta}
-\frac{1}{2}\,\tilde h_{\gamma\gamma}\delta_{\alpha\beta} +
\frac{1}{2}\partial_{\alpha}\partial_{\beta}\left(\Delta^{-1}
\tilde h_{\gamma\gamma}\right) ,
$$
where $\Delta^{-1}$ stands for the inverse of the Laplacian $\Delta
\equiv \delta^{\alpha\beta}\partial_{\alpha}\partial_{\beta}$. (The
existence of $\Delta^{-1}$ is guaranteed by the adopted boundary
conditions, which state that the perturbations $h_{\mu\nu}$ decrease
sufficiently fast in spatial infinity.) Then, the tensor part of the
equation (\ref{20c}) takes the simple form
\begin{equation}\label{33}
\left[\frac{1}{a^2}\hat h_{\alpha\beta}\right]_{,00} +
3\frac{\dot a}{a}\left[\frac{1}{a^2}
\hat h_{\alpha\beta}\right]_{,0} -
\frac{1}{a^2}\left[\frac{1}{a^2}
\hat h_{\alpha\beta}\right]_{,\gamma\gamma} = 0 \,.
\end{equation}
Being subject to the constraints $\hat h_{\alpha\alpha} = \hat
h_{\alpha\beta,\beta} = 0$, the variable $\hat h_{\alpha\beta}$
carries two physical degrees of freedom.

Before I proceed, let me draw your attention to the fact that the
scalar equation (\ref{28}) is identically satisfied if $H =
\mbox{const}$. This corresponds to the choice $a\propto e^{Ht}$. If
this is the case, the model (\ref{14}) reduces to GR with the
cosmological term---the model that carries only two physical degrees
of freedom. In what follows, I shall restrict my considerations to
the case $H \neq \mbox{const}$. Then, the scalar equation (\ref{28})
is rewritten as
\begin{equation}\label{34}
\left[ \frac{1}{a^2}\tilde h_{\alpha\alpha}\right]_{,00} +
\bigg(3H - 2\frac{\dot H}{H} + \frac{\ddot H}{\dot H} \bigg)
\left[ \frac{1}{a^2}\tilde h_{\alpha\alpha} \right]_{,0} -
\frac{1}{a^2}\left[ \frac{1}{a^2} \tilde
h_{\alpha\alpha}\right]_{,\beta\beta} = 0 \,.
\end{equation}
It is seen that the scalar mode of the geometric sigma model
(\ref{14}) is {\it massless}.

\section{Stability analysis}\label{Sec4}

In this section, I shall examine the stability of the vacuum solution
$h_{\mu\nu}=0$. Let me start with the tensor equation (\ref{33}).
First, I introduce the collective variable
$$
Q \equiv \frac{1}{a^2} \hat h_{\alpha\beta} \,.
$$
In terms of $Q$, the equation (\ref{33}) takes the compact form
\begin{equation}\label{35}
\ddot Q + 3 \frac{\dot a}{a} \dot Q -
\frac{1}{a^2} Q_{,\alpha\alpha} = 0 \,.
\end{equation}
The function $Q(x)$ is searched for in the form
\begin{equation}\label{36}
Q = {\rm Re} \int d^3 k \, q(k,t) \, e^{i\vec k\cdot \vec x} \,,
\end{equation}
whereupon the equation (\ref{35}) becomes
\begin{equation}\label{37}
\ddot q + 3 \frac{\dot a}{a} \dot q +
\frac{k^2}{a^2} q = 0 \,.
\end{equation}
The stability of the vacuum solution $q=0$ is examined by the
canonical analysis of the equation (\ref{37}). In the first step, I
notice that the equation (\ref{37}) is obtained from the Lagrangian
$$
{\cal L} = a \left( a^2 \dot q^2 - k^2 q^2 \right) .
$$
Indeed, it is easily verified that its variation leads to the
equation (\ref{37}). The corresponding Hamiltonian is
straightforwardly calculated to be
\begin{equation}\label{39}
{\cal H} = a \left( \frac{1}{4a^4}\, p^2 + k^2 q^2 \right) .
\end{equation}
It is seen that the Hamiltonian is positive for all the allowed
values of the parameter $a(t)$, and all the values of the wave vector
$\vec k$. Its minimum is located at $q=p=0$. For Hamiltonians with no
explicit time dependence, this would imply the stability of the
vacuum $q=p=0$. Indeed, owing to $\dot{\cal H} = \{{\cal H},{\cal
H}\} = 0$, the physical phase space trajectories coincide with the
orbits ${\cal H} = {\rm const}$. These orbits, on the other hand, are
closed curves around the vacuum $q=p=0$. As a consequence, the phase
space trajectory which is initially close to the vacuum continues to
be in the vicinity of the vacuum at all times.

\begin{figure}[htb]
\begin{center}
\includegraphics[height=2.5cm]{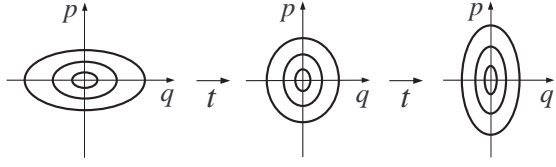}
\end{center}
\vspace*{-.5cm}
\caption{Time evolution of orbits $H={\rm const}$.\label{f1} }
\end{figure}

Unfortunately, the Hamiltonian (\ref{39}) depends on time through the
free parameter $a(t)$. As a consequence,
\begin{equation}\label{40}
\dot{\cal H} = \frac{\partial{\cal H}}{\partial t} +
\{ {\cal H},{\cal H} \} = \frac{\partial{\cal H}}{\partial t} \,,
\end{equation}
and the Hamiltonian is not conserved. Still, the stability of the
vacuum $q=p=0$ is not compromised. To see this, note that the orbits
${\cal H} = {\rm const.}$ remain to be closed curves around $q=p=0$,
only this time they evolve in time as shown in Fig. \ref{f1}.
It is seen from (\ref{40}) that the change of the Hamiltonian along
the phase space trajectory $(q(t),p(t))$ is the same as for the still
point $(q,p)={\rm const}$. Thus, the general solution of (\ref{40})
is the phase space trajectory which, at any time, touches the
respective orbit ${\cal H} = {\rm const}$. As these orbits are closed
curves around $q=p=0$, the phase space trajectories initially close
to $q=p=0$ remain to be close to $q=p=0$ at all times. In other
words, the vacuum solution $q=p=0$ is stable against small
perturbations. This holds true for any $a(t)$ and $\vec k$.

The stability of the scalar equation (\ref{34}) is examined
analogously. Using the notation
$$
Q\equiv \frac{1}{a^2}\tilde h_{\alpha\alpha}\,,
$$
the equation (\ref{34}) is rewritten as
\begin{equation}\label{40a}
\ddot Q +
\bigg(3H-2\frac{\dot H}{H}+\frac{\ddot H}{\dot H}\bigg)\dot Q -
\frac{1}{a^2} Q_{,\alpha\alpha} = 0 \,,
\end{equation}
and its Fourier transform, as defined by (\ref{36}), becomes
\begin{equation}\label{40b}
\ddot q +
\bigg(3H-2\frac{\dot H}{H} +
\frac{\ddot H}{\dot H}\bigg)\,\dot q +
\frac{k^2}{a^2}\,q = 0 \,.
\end{equation}
(This notation should not be confused with the same notation used in
the analysis of tensor modes.) It is easily checked that the equation
(\ref{40b}) follows from the Lagrangian
$$
{\cal L}\propto a\frac{\dot H}{H^2}\left(a^2 \dot q^2-k^2 q^2\right),
$$
or equivalently, from the Hamiltonian
\begin{equation}\label{40c}
{\cal H}\propto a\frac{\dot H}{H^2}\left(\frac{1}{4a^4}
\frac{H^4}{\dot H^2}\, p^2 + k^2 q^2 \right) .
\end{equation}
The proportionality sign reminds us that $\cal L$ and $\cal H$ are
determined only up to a multiplicative constant. If necessary, this
freedom can be used to correct the overall sign of the Hamiltonian.
As opposed to the Hamiltonian (\ref{39}), the Hamiltonian (\ref{40c})
is not necessarily positive. However, it is only the overall sign of
${\cal H}$ that can be negative. This implies that the vacuum $q=p=0$
can only be a maximum or a minimum, and never a saddle point. As a
consequence, the lines of constant ${\cal H}$ are closed curves
around $q=p=0$. From this point on, the stability analysis reduces to
that of the tensor mode, leading to the conclusion that the vacuum
$q=p=0$ is stable against scalar perturbations, too. This holds true
in a generic time interval, and for a generic choice of $a(t)$.

Let me note, however, that there can exist critical moments in the
evolution of scalar perturbations, where the stability may be lost.
As seen from (\ref{40c}), these are defined by $H = 0$ or $\dot H=0$.
In fact, the {\it stability discussed in this section is proven only
up to the presence of critical moments}. Such critical moments can be
found in many cosmological models, as will be demonstrated in Sec.
\ref{Sec5}. Some of them are benign, but some may cause divergent
behavior. It is important to realize that the construction of
geometric sigma models does not guarantee the absence of critical
moments. As a consequence, not every choice of the scale factor
$a(t)$ leads to an everywhere regular model. One should have in mind,
however, that the linear analysis considered in this paper is
inapplicable in the vicinity of critical moments. Indeed, the
perturbations grow big there, and the interaction terms can not be
neglected.

Before I close this section, let me mention once more that matter
fields have been excluded from the above analysis. As a consequence,
the proven stability may be compromised. Indeed, the target metric
$F(\phi)$ has been allowed to take negative values, and the potential
$V(\phi)$ to be unbounded from below. Then, the conventional coupling
to matter fields may lead to the appearance of ghosts and tachyons.
In this paper, however, I am examining the linear stability alone. It
will be demonstrated in the next section, that matter fields can not
appear in the linearized Einstein's and scalar field equations. Thus,
the established stability is not threatened. The stability of matter
perturbations themselves, on the other hand, is studied separately.

\section{Matter fields}\label{Sec4.5}

In this section, I shall examine how the presence of matter fields
influences the behavior of geometric sigma models. My starting point
is the action
\begin{equation}\label{60}
I = I_g + I_m \,,
\end{equation}
where $I_g$ is the geometric action (\ref{14}), and $I_m$ stands for
the action of matter fields. Customarily, the matter Lagrangian is
thought of as the Lagrangian of the standard model of elementary
particles, minimally coupled to gravity, and possibly, to the
inflaton field $\phi$. In what follows, the matter fields will
collectively be denoted by $\Omega$.

Let me first consider the simplest case characterized by the absence
of direct matter--inflaton couplings. Instead, the matter fields are
minimally coupled to the metric alone. This choice is justified by
the fact that minimal coupling to the metric already contains a
simple coupling to the inflaton itself. Indeed, it will shortly be
shown that matter fields do not compromise the gauge fixing procedure
of the preceding sections. This procedure leaves us with three
dynamical metric components, one of which is a reminiscent of the
original inflaton field. Of course, one can always change this simple
scenario by employing direct couplings to the inflaton field. It will
be shown later that most of the direct inflaton couplings preserve
the results of the preceding sections.

\subsection{Vacuum solution}\label{SubA}

In this subsection, I shall examine how the sigma model vacuum
(\ref{15}) is influenced by the presence of matter fields. Let me
start with the analysis of the matter field equations
\begin{equation}\label{61}
\frac{\delta I_m}{\delta\Omega} = 0 \,.
\end{equation}
Owing to the minimal coupling to the metric, these equations possess
vacuum solution that coincides with the well known vacuum of the
standard model of elementary particles. Indeed, the standard model
equations are trivially satisfied when matter fields take proper
constant values
$$
\Omega =\Omega_0 \,,
$$
and so are the equations (\ref{61}). This holds true for any value of
the metric that appears in the field equations. In particular, the
vacuum value of the stress-energy tensor
\begin{equation}\label{62}
T^{\mu\nu}_m \equiv
-\frac{2}{\sqrt{-g}}\,\frac{\delta I_m}{\delta g_{\mu\nu}}
\end{equation}
is zero, irrespectively of the presence of $g_{\mu\nu}$. Formally,
$$
\Omega = \Omega_0 \quad \Rightarrow \quad T^{\mu\nu}_m = 0
$$
for any $g_{\mu\nu}$. With this, the inflaton and Einsten's equations
take the form of the sigma model equations considered in the
preceding sections. Indeed, owing to the absence of the
matter-inflaton coupling, the inflaton equation is the same as that
of the sigma model,
$$
\frac{\delta I}{\delta \phi} = \frac{\delta I_g}{\delta \phi} = 0 \,.
$$
Einstein's equations, on the other hand, reduce to the sigma model
equations once the matter fields take their vacuum values,
$$
\Omega = \Omega_0 \quad \Rightarrow \quad
R^{\mu\nu}-\frac{1}{2}\,g^{\mu\nu}R = T^{\mu\nu}_{\phi} .
$$
It has already been shown in Sec. \ref{Sec2} that geometric sigma
model possesses the vacuum solution (\ref{15}). As a consequence, the
field equations that follow from the action (\ref{60}) are satisfied
by
\begin{equation}\label{64}
\Omega = \Omega_0 \,,\quad
\phi = t \,, \quad
g_{\mu\nu}=g^{(o)}_{\mu\nu}\,.
\end{equation}
This is the vacuum, or shall we say, the background solution of the
model (\ref{60}). It is seen that the presence of matter fields does
not compromise the sigma model vacuum of the preceding sections. This
result is summarized in the sentence
\begin{itemize}
\item {\it matter fields do not violate the sigma model vacuum}.
\end{itemize}
In what follows, I shall examine the stability of the vacuum
(\ref{64}) against its small perturbations. Owing to their smallness,
the dynamics of vacuum perturbations is governed by the linearized
field equations. Thus, it is the linear stability of the model
(\ref{60}) that will be examined in the next subsection.

\subsection{Stability analysis}\label{SubB}

The linear stability of the vacuum (\ref{64}) is examined by
inspecting the linearized field equations of the action (\ref{60}).
It is immediately seen that, after linearization, the inflaton and
Einstein's equations reduce to those of the geometric sigma model of
the preceding sections. Indeed, the stress-energy tensor
$T^{\mu\nu}_m$, being at least quadratic in perturbations of matter
fields, does not appear on the r.h.s. of the linearized Einstein's
equations. At the same time, the inflaton does not couple to matter
fields, at all. Hence, the linearized inflaton and metric equations
of motion remain unchanged by the inclusion of matter. They are
diffeomorphism invariant, so that the complete gauge fixing procedure
of Sec. \ref{Sec3} is still valid. In this gauge, the tensor and
scalar equations (\ref{33}) and (\ref{34}) are exactly what one
obtains from the linearized equations $\delta I/\delta\phi = 0$ and
$\delta I/\delta g_{\mu\nu} = 0$.

The generic linear stability of the sigma model vacuum has already
been established in the preceding section. What remains to be shown
is the linear stability of the matter field equations. To this end,
notice that the minimal coupling to matter fields implies the
validity of the {\it principle of equivalence}. Indeed, the matter
fields of the field equations $\delta I/\delta\Omega = 0$ are coupled
to at most first derivatives of the metric $g_{\mu\nu}$. In a local
inertial frame, the metric derivatives vanish, and the metric itself
becomes the Minkowski metric $\eta_{\mu\nu}$. This way, the matter
field equations take their special relativistic form---the standard
model of elementary particles in flat spacetime. The latter is known
to be stable against perturbations of its trivial vacuum. As a
consequence, the linearized matter field equations possess the stable
vacuum $\Omega = \Omega_0$. To summarize, the vacuum (\ref{64}) has a
generic linear stability against perturbations governed by the action
(\ref{60}). The dynamics of its geometric part remains the same as
found in Sec. \ref{Sec3}. Therefore,
\begin{itemize}
\item {\it the presence of matter fields does not violate the
established linear dynamics of geometric sigma models}.
\end{itemize}
What remains to be found is the dynamics of the linearized matter
field equations. There are three types of matter fields that appear
in the action: fermion fields, gauge fields and the Higgs. In the
linearized theory, fermion fields are governed by the Dirac equation,
gauge fields obey Maxwell equations, and the Higss is subject to the
Klein-Gordon equation. All of these are minimally coupled to the
external curved background.

{\bf Scalar fields.} The scalar $\varphi$, minimally coupled to the
external metric $g^{(o)}_{\mu\nu}$, obeys the Klein-Gordon equation
$(\Box - m^2)\varphi=0$. The scalar field mass is denoted by $m$, and
$\Box$ stands for the covariant d'Alembertian of the vacuum metric
$g^{(o)}_{\mu\nu}$. With $g^{(o)}_{\mu\nu}$ of the form (\ref{10}),
the Klein-Gordon equation becomes
$$
\ddot\varphi + 3\frac{\dot a}{a}\dot\varphi +
\left(m^2 - \frac{1}{a^2}\triangle\right)\varphi = 0 \,,
$$
where $\triangle \equiv \delta^{\alpha\beta} \partial_{\alpha}
\partial_{\beta}$. Now, one can use the Fourier decomposition
(\ref{36}) of the preceding section to rewrite the above equation in
the form
\begin{equation}\label{65}
\ddot q + 3\frac{\dot a}{a}\dot q +
\left(m^2 + \frac{k^2}{a^2}\right) q = 0 \,.
\end{equation}
This equation differs from the scalar perturbation equation
(\ref{40b}) not only by the presence of mass, but also by the
different friction coefficient. During the inflationary phase, when
Hubble parameter is approximately constant, the friction coefficient
of (\ref{40b}) is, in most cases, close to that of (\ref{65}). In
some cases, however, the scalar modes of the geometric sigma model
may have significantly higher friction, as will be demonstrated in
the next section. In such situations, the rapid expansion of the
Universe makes the inflaton decay much faster than scalars of the
matter Lagrangian.

{\bf Gauge fields.} In linear approximation, gauge fields obey the
equation $\nabla_{\mu}F^{\mu\nu}=0$, where $F_{\mu\nu} =
\partial_{\mu}A_{\nu} -\partial_{\nu}A_{\mu}$ is the linearized gauge
field strength, and $\nabla_{\mu}$ stands for the covariant
derivative. To simplify calculations, I shall work in the Coulomb
gauge. Then, the time component $A_0$ is constrained to be zero,
while the spatial components $A_{\alpha}$ satisfy
$$
\ddot A_{\alpha} + \frac{\dot a}{a}\dot A_{\alpha} -
\frac{1}{a^2}\triangle A_{\alpha} = 0 \,, \quad
\partial_{\alpha}A_{\alpha} = 0 \,.
$$
In terms of their Fourier components $q_{\alpha}$, the field
equations take the form
\begin{equation}\label{66}
\ddot q_{\alpha} + \frac{\dot a}{a} \dot q_{\alpha} +
\frac{k^2}{a^2} q_{\alpha} = 0 \,, \quad
k_{\alpha}q_{\alpha} = 0 \,.
\end{equation}
It is seen that the friction coefficient of (\ref{66}) is three times
smaller than that of (\ref{37}). Thus, in the expanding Universe, the
gauge fields outlast the excitations of the gravitational field.

{\bf Dirac field.} The evolution of Dirac field minimally coupled
to gravity is given by the equation
\begin{equation}\label{67}
\left(i\gamma^k \nabla_k - m\right)\psi =0 \,,
\end{equation}
where $\gamma^k$ are Dirac gamma matrices ($\{\gamma^i,\gamma^j
\}=-2\eta^{ij}$), and $\nabla_k$ stands for the covariant derivative
$$
\nabla_k \psi = e_k{}^{\mu}\left(\partial_{\mu} +
\omega_{\mu} \right) \psi \,.
$$
The gravitational variables that enter this equation are the tetrad
$e^k{}_{\mu}$, and the spin connection $\omega^{ij}{}_{\mu}$. The
tetrad $e^k{}_{\mu}$ is the inverse of $e_k{}^{\mu}$, while
$\omega^{ij}{}_{\mu}$ are the components of $\omega_{\mu}$ in the
basis of Lorentz generators $\sigma_{ij} \equiv \frac{1}{4}
\left[\gamma_i,\gamma_j \right]$,
$$
\omega_{\mu} = \frac{1}{2}\omega^{ij}{}_{\mu}\sigma_{ij}\,.
$$
The metric $g_{\mu\nu}$ and the connection
$\Gamma^{\lambda}{}_{\mu\nu}$, which are used in this paper, are
related to $e^k{}_{\mu}$ and $\omega^{ij}{}_{\mu}$ through the
equations \cite{34}
\begin{equation}\label{68}
g_{\mu\nu} = \eta_{ij} e^i{}_{\mu} e^j{}_{\nu} ,\quad
e^i{}_{\mu,\nu} + \omega^i{}_{j\nu} e^j{}_{\mu} -
\Gamma^{\lambda}{}_{\mu\nu}e^i{}_{\lambda} = 0 \,.
\end{equation}
The first equation is the very definition of the orthonormal tetrad,
while the second represents the metricity condition. It is seen that,
given the metric $g_{\mu\nu}$, the tetrad is determined only up to
the local Lorentz rotations. These, however, are a symmetry of the
Dirac equation (\ref{67}), and can be gauge fixed. Thus, starting
with the background metric $g^{(o)}_{\mu\nu}$, the simplest solution
for the background tetrad is found to have the diagonal form
$$
e^0{}_0 = 1 \,, \quad
e^1{}_1 = e^2{}_2 = e^3{}_3 = a \,.
$$
Then, the second equation (\ref{68}) yields the background value of
the spin connection. The only non-zero components turn out to be
$$
\omega^{b0}{}_{\alpha} = -\omega^{0b}{}_{\alpha} =
\dot a \delta^b_{\alpha} \,.
$$
With the known background values of the tetrad and spin connection,
the Dirac equation (\ref{67}) is rewritten as
$$
\dot\psi+\frac{1}{a}\gamma^0\gamma^{\alpha}\psi_{,\alpha} +
\frac{1}{2}\left(3\frac{\dot a}{a}+2im\gamma^0\right)\psi = 0 \,.
$$
Its Fourier expansion then yields
\begin{equation}\label{69}
\dot q + \frac{1}{2}\left[3\frac{\dot a}{a} + 2i\gamma^0
\left(m+\frac{1}{a}\vec\gamma\cdot\vec k\right)\right]q = 0 \,,
\end{equation}
where $q(\vec k,t)$ are defined by
$$
\psi(x) = \int d^3 k\,q(\vec k,t)\,e^{i\vec k\cdot\vec x} \,.
$$
Using the notation
$$
q=\left(
\begin{array}{c}
u \\ v \\
\end{array}\right) , \quad
u=\left(
\begin{array}{c}
u_+ \\ u_- \\
\end{array}\right), \quad
v=\left(
\begin{array}{c}
v_+ \\ v_- \\
\end{array}\right),
$$
the $4$-component equation (\ref{69}) is rewritten as a pair of
$2$-component equations. The first is the constraint equation
$$
v=\frac{i}{m}\left[\dot u + \frac{1}{2a}\left(
3\dot a - 2i\vec \sigma \cdot \vec k \right) u \right] ,
$$
which tells us that $v$ carries no degrees of freedom. The second is
the dynamical equation
$$
\ddot u + 3H\dot u + \left[ m^2 +
\frac{\vec k}{a}\cdot\left(\frac{\vec k}{a} +
iH\vec\sigma\right) + \frac{3}{4} \left( 2\dot H+3H^2 \right)
\right] u = 0 \,.
$$
If the coordinates are rotated so that $\vec k$ is directed along the
$z$-axis, the above equation turns into a system of two one-component
equations
\begin{equation}\label{70}
\ddot u_{\pm} + 3H\dot u_{\pm} + \left[ m^2 +
\frac{k}{a}\left(\frac{k}{a} \pm iH\right) +
\frac{3}{4} \left( 2\dot H+3H^2 \right)\right] u_{\pm} = 0 \,.
\end{equation}
It is seen that the friction coefficient coincides with that of the
scalar field, whereas the mass term is different. In the early stage
of inflation, when $a$ is still very small, the inequality $k\gg aH$
holds for a wide range of $k$. In this regime, the mass term is close
to that of the scalar field. At the end of inflation, however, the
scale factor is large, so that most $k$ satisfy $k\ll aH$. Then, the
mass term takes the form $m^2+(3H/2)^2$. As the inflationary value of
$H$ is typically much larger than masses of elementary fermions, the
equation (\ref{70}) is practically independent of $m$. This means
that the production and propagation of elementary fermions are the
same for all the fermion species.

\subsection{Matter-inflaton coupling}\label{SubC}

So far, the considered matter fields have been assumed to couple to
the metric alone. It has been shown that the results of the previous
sections are not compromised by the inclusion of such matter. In this
subsection, I shall discuss direct matter--inflaton couplings.

Let me start with the verification of the vacuum solution (\ref{64}).
It is immediately seen that interaction terms which are at least
quadratic in perturbations of matter fields do not compromise the
vacuum (\ref{64}). Indeed, their contribution to the field equations
disappears when $\psi=\psi_0$, thereby making (\ref{64}) a valid
vacuum solution. The sigma model vacuum (\ref{15}), on the other
hand, satisfies the field equations even when the couplings are
linear in perturbations of matter fields. Thus,
\begin{itemize}
\item {\it typical matter--inflaton couplings preserve the sigma
model vacuum}.
\end{itemize}
The examples of possible matter--inflaton couplings are
$$
f(\phi)g(\chi) \,,\quad
f(\phi){\bar\psi}\psi \,,\quad
f(\phi)F_{\mu\nu}F^{\mu\nu}
$$
and so on. It is seen that symmetries of the standard model
Lagrangian restrict most of these interaction terms to be at least
quadratic in matter fields. This is fortunate because such
matter-inflaton couplings preserve the results of the preceding
sections. Indeed, the linearized Einstein's and inflaton equations
remain the same as those obtained in the absence of matter fields.
Thus,
\begin{itemize}
\item {\it the analysis of the preceding sections is not compromised
by the presence of interaction terms which are at least quadratic in
matter fields}.
\end{itemize}
What does change, however, is the form of the linearized matter field
equations. As an example, let me consider Dirac field with the
inflaton coupling of the form $f(\phi){\bar\psi}\psi$. Then, the
linearized equations (\ref{70}) take the form
\begin{equation}\label{71}
\ddot u_{\pm} + A\dot u_{\pm} + \left[ M^2 +
\frac{k}{a}\left(\frac{k}{a} \pm iB\right) + \frac{3}{4}H\left(
2C+3H\right)\right] u_{\pm} = 0 \,,
\end{equation}
where $M(t)\equiv  m-f(t)$, and
$$
A \equiv 3H - \frac{\dot M}{M} \,,\quad
B \equiv H +  \frac{\dot M}{M} \,,\quad
C \equiv \frac{\dot H}{H} - \frac{\dot M}{M} \,.
$$
It is seen that the equation (\ref{71}) has singular points defined
by $M(t)=0$. The easiest way to get rid of these is to restrain the
couplings to satisfy $f(\phi) < m$. Such is, for example, the
coupling $f=\xi\phi^2$ when $\xi<0$. The simplest choice
$f=\lambda\phi$ makes the perturbations $u_{\pm}$ diverge at
$t=m/\lambda$. However, this coupling is perfectly acceptable in
cosmological models whose initial singularity is located at $t\geq
m/\lambda$. For example, if the Universe is born at $t=0$, the
coupling $f=\lambda\phi$ with $\lambda<0$ yields an everywhere
regular dynamics.

Before I close this section, let me note that matter-inflaton
coupling can significantly modify the dynamics of matter fields. In
particular, the particle production rate can be increased. For
example, the coupling $f=\lambda\phi$ with $\lambda<0$ diminishes the
value of the friction coefficient to
$$
A=3H+\frac{\lambda}{m-\lambda t} \,.
$$
If $|\lambda|$ is large enough, there is a time interval in which the
friction becomes negative. During that time, the initial matter
perturbations keep growing, despite the rapid expansion of the
Universe. Obviously, this makes the production of particles more
effective. Unfortunately, the negative friction in this example does
not last long. Compared with the typical inflationary period, it is
more than one hundred times shorter. Still, it is always possible to
adjust the matter-inflaton coupling to obtain the desired particle
production rate.

\section{Examples}\label{Sec5}

In this section, I shall analyze three specific choices of the
Universe dynamics. In the first, a toy model is used for the
demonstration of how the described procedure works in practice. In
the second, I propose an inflationary model close to the standard
model of the Universe. The third is a model of the bouncing Universe.
Neither of these models considers matter fields. They are solely used
for the demonstration of how a dark energy action is associated with
a chosen background geometry of the Universe. Hopefully, the right
choice of the scale factor $a(t)$ will ultimately lead to a realistic
cosmological model.

\subsection{Toy model}\label{Sub1}

\begin{figure}[htb]
\begin{center}
\includegraphics[height=4cm]{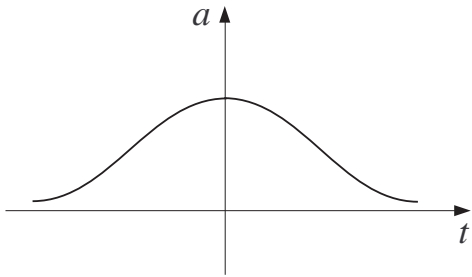}
\end{center}
\vspace*{-.5cm}
\caption{Toy model.\label{f2} }
\end{figure}

Let me consider a homogeneous, isotropic and spatially flat geometry
(\ref{10}) with the scale factor of the simple form
\begin{equation}\label{41}
a(t) = \frac{1}{\cosh \omega t} \,.
\end{equation}
Its graph is displayed in Fig. \ref{f2}. It describes an ever
existing Universe, whose inflationary epoch begins at infinite past
and lasts
infinitely many $e$-folds. The  exit from inflation happens at
$t\approx -1/\omega$. The constant $\omega$ is a free parameter of
the model.

\begin{figure}[htb]
\begin{center}
\includegraphics[height=4cm]{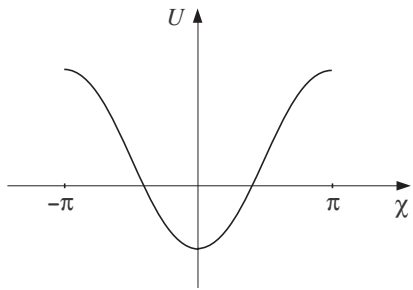}
\end{center}
\vspace*{-.5cm}
\caption{Potential function $U(\chi)$.\label{f3} }
\end{figure}

\begin{figure}[htb]
\begin{center}
\includegraphics[height=4.5cm]{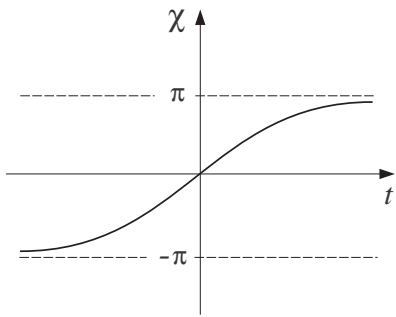}
\end{center}
\vspace*{-.5cm}
\caption{Soliton solution.\label{f4} }
\end{figure}

The scale factor (\ref{41}) is a solution of the sigma model
(\ref{14}) in which the potential $V(\phi)$ and the target metric
$F(\phi)$ are calculated from (\ref{12}) and (\ref{13}). This
procedure straightforwardly leads to
\begin{equation}\label{42}
F(\phi) = \frac{2\omega^2}{\cosh^2 \omega\phi} \,, \quad
V(\phi) = \frac{6\sinh^2\omega\phi-2}{\cosh^2 \omega\phi}
\,\omega^2 .
\end{equation}
With $F(\phi)$ and $V(\phi)$ defined by (\ref{42}), the action
functional (\ref{14}) has the solution (\ref{15}), in which
$g^{(o)}_{\mu\nu}$ is defined by (\ref{10}) and (\ref{41}). The model
can further be simplified by a suitable redefinition of the scalar
field. Specifically, the redefinition $\phi\to\chi(\phi)$ of the form
\begin{equation}\label{43}
\chi \equiv 2 \arctan \left( \sinh\omega\phi \right)
\end{equation}
brings the action (\ref{14}) to the form
$$
I_g = \frac{1}{2\kappa}\int d^4x\sqrt{-g}\left[ R -
\frac{1}{2}\chi^{,\mu}\chi_{,\mu} - U(\chi) \right] .
$$
The new potential $U(\chi) = V(\phi(\chi))$ reads
$$
U(\chi) = 2\,\omega^2 \left( 4\sin^2\frac{\chi}{2} - 1 \right) .
$$
In terms of $\chi$, the nontrivial vacuum solution $\phi = t$ becomes
\begin{equation}\label{46}
\chi(t) = 2 \arctan \left( \sinh\omega t \right) ,
\end{equation}
while the metric $g_{\mu\nu} = g^{(o)}_{\mu\nu}$ remains the same.
The graphs of the potential $U(\chi)$, and the solution (\ref{46})
are displayed in Figs. \ref{f3} and \ref{f4}.
It is seen that the potential $U(\chi)$ is a periodic function of
$\chi$, with the period $2\pi$. Thus, the scalar field $\chi$ lives
on a circle. The soliton solution (\ref{46}) is one-to-one mapping
$R^1\to S^1$. As for the trivial solutions, the theory accommodates
two of them. The first is given by $\chi=0$, and Anti-de Sitter
metric. The second has $\chi=\pm\,\pi$, and the metric is de Sitter.
Only the first solution is stable, because $\chi=0$ is the minimum of
the potential $U(\chi)$, and Anti-de Sitter metric is a stable
solution of the corresponding geometric equation \cite{33}. The
unstable solution $\chi = \pm\,\pi$ is the limiting case of the
stable soliton solution when $t\to\pm\,\infty$. It seems as if stable
soliton is asymptotically unstable. However, there is no
contradiction in this unusual situation. This is because the
stability analysis of Sec. \ref{Sec4} deals with the infinitesimal
perturbations, which preserve the monotonous character of the soliton
solution. As a consequence, the scalar $\chi$ can be gauged away.
This is not the case with the trivial vacuum $\chi = \pm\,\pi$. No
matter how small the perturbations of this vacuum are, they can not
be gauged away. This is why the stability of $\chi = \pm\,\pi$ can
not be a substitute for the asymptotic stability of the soliton
vacuum.

Before I close this subsection, let me say something about
perturbations of the background solution (\ref{41}). It is
immediately seen that $t=0$ is the only critical point of this model.
In this point, the coefficients of the tensor equation (\ref{37}) are
regular, but the friction coefficient of the scalar equation
(\ref{40b}) diverges as $-2/t$. Luckily, this singularity turns out
not to be harmful. Indeed, a careful analysis shows that scalar
perturbations formed in the past regularly pass the critical point
$t=0$.

Finally, let me calculate the friction coefficient of the scalar mode
of this model. A simple calculation shows that it approaches the
value $5\omega$ when $t\to -\infty$. In the same epoch, the friction
coefficient of matter scalars (which obey the equation (\ref{65}))
has the value $3\omega$. Thus, during rapid expansion of the
Universe, the inflaton decays faster than matter scalars. Moreover,
if the model is defined by
$$
a(t) = \left(\cosh\omega t\right)^{-\frac{1}{n}} \,,
$$
the ratio of the two friction coefficients becomes $1+2n/3$. When
$n\gg 1$, the inflaton decay rate becomes much larger than that of
matter scalars.

\subsection{Inflationary Universe}\label{Sub2}

In the second example, I shall examine the scale factor of the form
\begin{equation}\label{47}
a(t) = \Big[ 1+ \tanh (8\omega t)  \Big]
\ln \left( 1+ \exp{\frac{\omega t - 4}{4}} \right) .
\end{equation}
As seen from its graph in Fig. 5, it mimics the standard model of the
Universe. Indeed, all the expected phases of the cosmological
\begin{figure}[htb]
\begin{center}
\includegraphics[height=3.7cm]{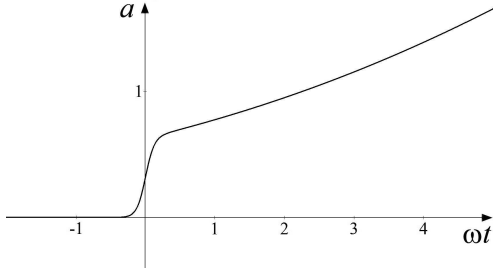}
\end{center}
\vspace*{-.5cm}
\caption{Inflationary Universe.\label{f5} }
\end{figure}
evolution are there. The inflationary epoch begins at infinite past,
and lasts infinitely many $e$-folds. The exit from inflation happens
at $\omega t \approx 0$, when the early acceleration stops. The
Universe continues to expand slowly, until it reaches the moment when
the late time acceleration begins. The present epoch is located at
$\omega t \approx 7.7$. I shall demonstrate later how the parameter
$\omega$ and the present time $t_{\rm now}$ are calculated from the
known values of the Hubble and deceleration parameters.

The action functional whose vacuum solution is defined by (\ref{47})
has the form (\ref{14}), with $F(\phi)$ and $V(\phi)$ calculated from
(\ref{12}) and (\ref{13}). A straightforward procedure leads to the
cumbersome expressions which I choose not to display here. Instead,
\begin{figure}[htb]
\begin{center}
\includegraphics[height=3.5cm]{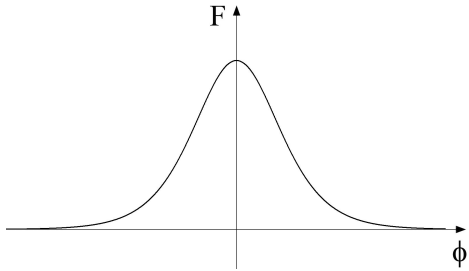}
\end{center}
\vspace*{-.5cm}
\caption{Target metric $F(\phi)$.\label{f6} }
\end{figure}
their graphs are presented. The action (\ref{14}), with $F(\phi)$ and
$V(\phi)$ depicted in Figs. \ref{f6} and \ref{f7}, has the soliton
solution $\phi = t$.
\begin{figure}[htb]
\begin{center}
\includegraphics[height=3.5cm]{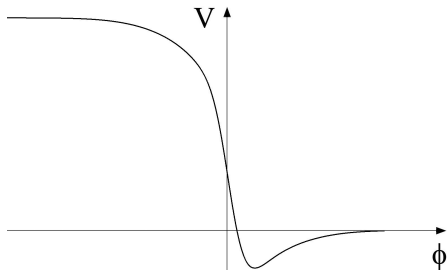}
\end{center}
\vspace*{-.5cm}
\caption{Potential $V(\phi)$.\label{f7} }
\end{figure}

The parameter $\omega$ can be determined from the known values of the
Hubble and deceleration parameters. These are defined as
$$
H \equiv \frac{\dot a}{a} \,, \qquad
q \equiv -\frac{\ddot a}{a H^2} \,.
$$
With the scale factor given by (\ref{47}), the parameters $H/\omega$
and $q$ depend on $\omega$ and $t$ only through the combination
\begin{figure}[htb]
\begin{center}
\includegraphics[height=4cm]{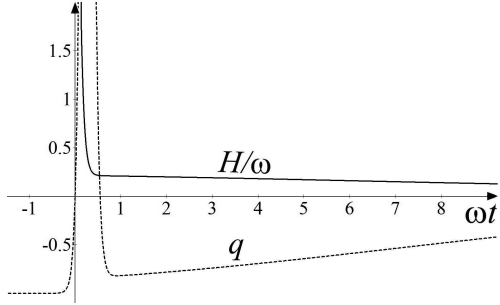}
\end{center}
\vspace*{-.5cm}
\caption{Hubble and deceleration parameters.\label{f8} }
\end{figure}
$\omega t$. The direct calculation yields the functions whose graphs
are displayed in Fig. \ref{f8}. Using these, one easily finds that
the astronomically observed values
$$
H = 0.075 \ \,{\rm Gyr}^{-1} \,, \qquad  q = -\,0.5
$$
imply $\omega t = 7.68$ and $H/\omega = 0.14$. Thus,
$$
\omega = 0.53 \ \,{\rm Gyr}^{-1} \,, \qquad
t_{\rm now} = 14.55 \ \,{\rm Gyr} \,.
$$
The obtained value of $t_{\rm now}$ is the time coordinate measured
from the arbitrarily chosen origin $t=0$. This is not what one would
like to have. Instead, the present epoch should be measured relative
to a physically significant moment in the history of the Universe.
This can not be the initial singularity, as our model does not have
one. Instead, my choice is the end of inflation. The end of inflation
$t_{\rm inf}$ is naturally defined as the moment when the early
acceleration stops. It is seen from the graph in Fig. \ref{f5} that
$\omega t_{\rm inf} = 0.003$. This leads to
\begin{equation}\label{51}
t_{\rm now} - t_{\rm inf} = 14.54 \ \,{\rm Gyr} \,.
\end{equation}
The time interval (\ref{51}) is a substitute for what is commonly
called the age of the Universe.

So far, I have discussed everywhere regular cosmological models.
However, these can easily be modified to become models with the
initial singularity. For example, the model under consideration can be
redefined by replacing its scale factor $a(t)$ with
$$
\tilde a(t) = a(t) - a(t_0) \,.
$$
The new scale factor describes a Universe which is born at $t=t_0$,
and lives regularly ever after. As opposed to the inflationary epoch
of everywhere regular model (\ref{47}), the inflationary epoch of the
new model lasts a finite number of $e$-folds. This number can be made
arbitrarily large by letting $t_0 \to -\infty$. It can also be shown
that the new model has no singularities other than $t=t_0.$

Finally, let me briefly discuss the propagation of small
perturbations. It is seen from Figs. \ref{f5} and \ref{f6} that
neither the Hubble parameter $H$, nor the target metric $F$ have
zeros. As a consequence, the model under consideration has no
critical moments. This means that both, tensor and scalar,
perturbations have everywhere regular dynamics. In particular, their
friction coefficients are everywhere finite. While tensor
fluctuations have always positive friction, the friction of the
scalar fluctuations can become negative. Indeed, a straightforward
analysis shows that, after the inflation, the scalar friction
abruptly drops and becomes negative. The period of negative friction
does not last long. When the late time acceleration begins, it
already has a small positive value which gradually approaches zero as
$t\to\infty$. In the next subsection, I shall present the example of
a model whose scalar fluctuations have critical moments.

\subsection{Bouncing Universe}\label{Sub3}

\begin{figure}[htb]
\begin{center}
\includegraphics[height=4cm]{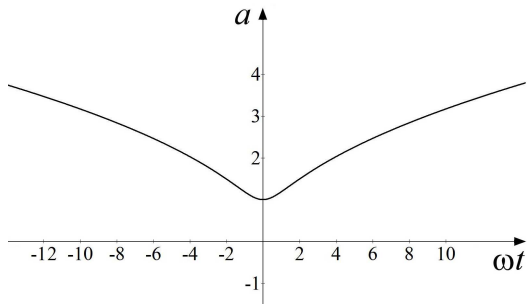}
\end{center}
\vspace*{-.5cm}
\caption{Bouncing Universe.\label{f9} }
\end{figure}

In this example, I shall consider the scale factor of the form
\begin{equation}\label{52}
a(t) = \left( 1+\omega^2 t^2 \right)^{\frac{1}{4}} .
\end{equation}
Its graph is displayed in Fig. \ref{f9}. It defines a bouncing
Universe which begins in the infinite past, slowly shrinks to its
minimal size, and then bounces to an expanding phase. The action
functional whose vacuum solution is defined by (\ref{52}) has the
form (\ref{14}), with $F(\phi)$ and $V(\phi)$ calculated from
(\ref{12}) and (\ref{13}). A straightforward procedure leads to
\begin{equation}\label{53}
F(\phi)=\omega^2 \frac{\omega^2\phi^2 -1}
{\left(\omega^2\phi^2 +1\right)^2} \,,               \qquad
V(\phi)=\omega^2\frac{\omega^2\phi^2 +2}
{2\left(\omega^2\phi^2 +1\right)^2} \,.
\end{equation}
The action (\ref{14}), with $F(\phi)$ and $V(\phi)$ defined by
(\ref{53}), has the spatially homogeneous and isotropic solution
$$
\phi(t) = t \,, \quad
ds^2=-dt^2+\sqrt{1+\omega^2 t^2}
\left(dx^2+dy^2+dz^2\right).
$$
It is seen that the background spacetime is flat in the infinite past
and future. Indeed, all the curvature invariants are shown to fall
off as $1/t^2$ or faster as $|t|\to\infty$. Thus, the Universe in
this example evolves out of the flat spacetime.

How do metric perturbations propagate in this background? A simple
analysis shows that tensor perturbations are everywhere regular,
whereas scalar ones have critical points. It is straightforward to
verify that there are three critical points: the two zeroes of $\dot
H$, located in $t=\pm 1/\omega$, and the zero of $H$, located in
$t=0$. These are the singularities of the friction coefficient in
(\ref{40b}). Its behavior in the vicinity of the critical points is
given by
$$
3H-2\frac{\dot H}{H}+\frac{\ddot H}{\dot H} \sim
\left\{
\begin{array}{cl}
\ds -\frac{2}{t} & \ds \mbox{in the vicinity of }\,t=0\,,\\ [0.7em]
\ds \frac{1}{t\pm 1/\omega} &
\ds \mbox{in the vicinity of }\, t = \mp\, 1/\omega \,.
\end{array}
\right.
$$
As has already been mentioned, the singularity of the form $-2/t$
does not violate the regular propagation of scalar perturbations.
However, it is not the case with the singularities $t=\pm 1/\omega$.
These are shown to act as barriers to the scalar perturbations coming
from the past. Indeed, the scalar perturbations go to infinity as
they approach $t=\pm 1/\omega$. This is an undesirable property,
which calls for the rejection of the model. Still, one should be
aware of the fact that our linear analysis makes no sense if the
fields are too strong. In such situations, the interaction terms
should be taken into account. Hopefully, this could cure the
singularity problem.

\section{Recapitulation}\label{Sec6}

The purpose of this work has been to demonstrate how an arbitrarily
chosen background of the Universe can be made a solution of a simple
model. To this end, I made use of the concept of geometric sigma
models. These models possess two distinctive features. The first is
that any metric can be made a solution of a particular geometric
sigma model. The second ensures that the complete matter content can
be gauged away. In this paper, a geometric sigma model is associated
with an arbitrary homogeneous, isotropic and spatially flat geometry.
In its simplest form, the model describes one scalar field in
interaction with Einstein's gravity. It possesses the vacuum solution
$\phi=t$, $g_{\mu\nu} = g_{\mu\nu}^{(o)}$. It is important to
emphasize that, while the background metric $g_{\mu\nu}^{(o)}$ can be
chosen arbitrarily, the physics of its small perturbations can not.
In fact, the role of the background metrics $g_{\mu\nu}^{(o)}$ is to
parametrize the class of models presented in this paper. This way,
the search for a viable cosmological model reduces to the proper
choice of the background metric.

The present work begins with the recapitulation of the concept of
geometric sigma models. The construction of the generic model is
presented, and subsequently applied to the homogeneous, isotropic and
spatially flat geometry of the Universe. Then, the dynamics of small,
localized perturbations of the vacuum is examined. It is demonstrated
how all but three degrees of freedom can be gauged away. The
classical stability of the gauge fixed linearized theory is proven in
Sec. \ref{Sec4}. This is done by direct calculation, as stability of
the vacuum solution is not guaranteed by the very construction of
geometric sigma models. The vacuum stability against matter
fluctuations is considered in Sec. \ref{Sec4.5}. It is shown that the
inclusion of matter fields does not compromise the results of the
preceding sections. The analysis of the stability against matter
perturbations concludes the general considerations of the paper.

The rest of the paper is devoted to examples. The first is a toy
model used to demonstrate how the method works in practice. The
second is the example of an inflationary Universe, and the third is a
bouncing model. The corresponding action functionals are constructed
along the lines described in Sec. \ref{Sec2}. The obtained target
metric $F(\phi)$, and the potential $V(\phi)$ are presented through
their graphs. The graphical method is also used for numerical
calculations. In particular, it is demonstrated how the parameters of
the model are calculated from the known values of the Hubble and
deceleration parameters.

I have also given a brief insight into the propagation of small
perturbations. It is argued that the generic linear stability proven in
Sec. \ref{Sec4} fails in the vicinity of critical moments $\dot H = 0$.
There, the scalar perturbations diverge, so that a consistent
stability analysis must go beyond the linear approximation. A more
elaborate analysis of this topic is left for the future
investigation. In particular, the nonperturbative analysis near the
critical moments, and the study of more complex geometric sigma
models is planed.

At the end, let me point out that the procedure described in this
paper misses an important ingredient. It concerns the back reaction
of quantum vacuum fluctuations on the background geometry. Indeed,
the only matter that geometric sigma models of Sec. \ref{Sec2} deal
with is a specific dark energy of purely geometric origin. Although
it successfully generates any desirable geometry of the Universe, the
influence of quantum fluctuations of ordinary matter has not been
taken into account. Instead, it has been demonstrated in Sec.
\ref{Sec4.5} that matter fields preserve the classical linear
stability established in Sec. \ref{Sec4}. The completion of the
incomplete cosmology presented in this paper requires the inclusion
of quantum fluctuations of both geometry and matter. Until then, it
is comforting to know that dark energy is commonly believed to
dominate all other forms of matter in the Universe. Owing to this,
the predictions of this work may not be far from realistic, after
all.

\begin{acknowledgments}
This work is supported by the Serbian Ministry of Education, Science
and Technological Development, under Contract No. $171031$.
\end{acknowledgments}

\end{document}